\RequirePackage{amsmath}
\documentclass[structabstract]{aa}

\usepackage[utf8]{inputenc}

\usepackage{amsmath}

\usepackage{calc}

\usepackage{longtable}
\usepackage{booktabs}
\usepackage{array}

\usepackage{color}

\usepackage[pdftex]{graphicx}
\usepackage{subfigure}

\usepackage{txfonts}
\usepackage{ae}

\usepackage{natbib}
\bibpunct{(}{)}{;}{a}{}{,} % to follow the A&A style

\usepackage{units}

\usepackage{textcomp}
\usepackage{gensymb}

\usepackage[colorlinks=true,linkcolor=blue,citecolor=blue,bookmarks]{hyperref}

\begin{document}

\title{Morphometric analysis in gamma-ray astronomy\\
  using Minkowski functionals:}
\subtitle{III. Sensitivity increase via a refined structure quantification}

\author{M.~A.~Klatt\inst{\ref{kit},\ref{theo1},\ref{ecap}}
  \thanks{\email{michael.klatt@kit.edu}}
  \and
  K.~Mecke\inst{\ref{theo1},\ref{ecap}}
}

\titlerunning{Morphometric analysis in gamma-ray astronomy}
\authorrunning{M.~A.~Klatt \and K. Mecke}

\institute{
  Karlsruhe Institute of Technology (KIT), Institute of Stochastics, Englerstr. 2, 76131 Karlsruhe, Germany\label{kit}
  \and
  Institut für Theoretische Physik,
  Universität Erlangen-Nürnberg,
  Staudtstr. 7,
  91058 Erlangen, Germany\label{theo1}
  \and
  Erlangen Centre for Astroparticle Physics,
  Universität Erlangen-Nürnberg,
  Erwin-Rommel-Str. 1,
  91058 Erlangen, Germany\label{ecap}
}

\date{Received \dots / Accepted \dots}

% \abstract{}{}{}{}{}
% 5 {} token are mandatory

\abstract
{}
{ We pursue a novel morphometric analysis to detect sources in very-high-energy gamma-ray counts maps by structural deviations from the background noise without assuming any prior knowledge about potential sources.
  The rich and complex structure of the background noise is characterized by Minkowski functionals from integral geometry.
  By extracting more information out of the same data, we aim for an increased sensitivity. 
}
{ In the first two papers, we derived accurate estimates of the joint distribution of all Minkowski functionals.
  Here, we use this detailed structure characterization to detect structural deviations from the background noise in a null hypothesis test.
  We compare the analysis of the same simulated data with either a single or all Minkowski functionals.
}
{ The joint structure quantification can detect formerly undetected sources.
  We show how the additional shape information leads to the increase in sensitivity.
  We explain the very unique concepts and possibilites of our analysis compared to a standard counting method in gamma-ray astronomy,
  and we present in an outlook further improvements especially for the detection of diffuse background radiation and generalizations of our technique.
}
{}


\maketitle

\section{Morphometric source detection in gamma-ray astronomy}
\label{Introduction}

The morphometric analysis quantifies the complex structural information that is contained in the background noise in ground-based Very-High Energy (VHE) gamma-ray astronomy.
We thus quantify the shape of sky maps without any assumption about potential sources~\citep{KlattEtAl2012, GoeringKlattEtAl2013}.

The Minkowski functionals from integral geometry can comprehensively and robustly quantify the complex shape provided by spatial data~\citep{SchneiderWeil2008, MantzJacobsMecke2008, SchroederTurketal:2010jom, SchroederTurketal2010AdvMater, SchroederTurkEtAl2013NewJPhys}. 
They allow for an efficient data analysis and a sensitive hypothesis test.
Because of their versatility, they cannot only check for specific structures or arrangements but can detect any structural deviation from the expected background.
In other words, by characterizing the shape of a noisy sky-map more information can be taken out of the same data without assuming prior knowledge about the source.

This is in contrast to the common null hypothesis test by \citet{LiMa1983} that only uses the total number of counts but no further geometric information, which might be especially valuable for the analysis of 
extended sources or diffuse VHE emissions~\cite{rxj1713, velajr, galcen}.
It is also in contrast to full likelihood fits of models to the measured data of high-energy gamma-ray telescopes~\citep{egretlike, fermilat}, which strongly depend on the model and on the a-priori knowledge about the sources.

The methods and concepts in this article are in principle applicable to any random field and any spatial data to detect inhomogeneities or other structural deviations.
It could, for example, be interesting for medical data sets, e.g., in tumor recognition~\citep{CanutoEtAl2009, LarkinEtAl2014, ArfelliEtAl2000, MichelEtAl2013},
for geospatial data and raster data in earth science~\citep{StonebrakerEtAl1993},
in image and video analysis, where a fast analysis for the detection of objects is needed~\citep{BorgEtAl2005, YilmazEtAl2006, QuastKaup2011},
or in the related field of pattern recognition~\citep{JainEtAl2000, TheodoridisKoutroumbas2009}.
However, the technique is especially interesting for very high energy (VHE) gamma-ray astronomy,
where faint extended signals are overlaid by strong background noise~\citep{BuckleyEtAl2008}.
Especially, for short observation times and low statistics, that is, when an increase in sensitivity is most needed,
the advantage of additional structural information should be most effective: the excess in the number of counts
might not be significant because of the strong fluctuations of a Poisson distribution relative to the small mean value.
However, the improbable spatial arrangement of the fluctuations can eventually lead to the detection of the source.

In astronomy, the Minkowski functionals are already used 
as probes of non-Gaussianity in the cosmic microwave background
\citep{Schmalzing1999,Novikov:441276,Gay2012,Ducout2013,NovaesEtAl2014},
to characterize nuclear matter in supernova explosions~\citep{SchuetrumpfKlattEtAl2013, SchuetrumpfKlattEtAl2014},
and to investigate the large-scale structure of the
universe~\citep{MeckeBuchertWagner1994, Colombi2000, KerscherEtAl2001, KerscherMecke2001, wiegand_direct_2014}.

In the first two papers of this series, we introduced the morphometric analysis to gamma-ray astronomy and derived accurate estimates of the structure distribution of the background.
Here, we apply this refined shape analysis to simulated data and demonstrate how the additional geometric information can lead to a strong increase in sensitivity.

% Outline

In Sec.~\ref{sec_gamma_short_overview}, we shortly summarize the most important steps of the morphometric analysis.
In Sec.~\ref{sec_gamma_sensitivity}, we demonstrate the increase in sensitivity via a refined structure quantification.
By analyzing the same data simultaneously by all three Minkowski functionals instead of a simple structure characterization, the compatibility with the background structure can decrease by 14 orders of magnitude,
that is, the probability to find such a fluctuation in the background is $10^{-19}$ instead of $10^{-5}$.
Formerly undetected sources can thus eventually be detected, which depends of course on the shape of the source whether there is a structural deviation or not.

The morphometric analysis is then compared in Section~\ref{sec_gamma_compare_lima} to a standard null hypothesis test in gamma-ray astronomy.
The comparison depends both on the shape of the source and on the experimental details.
An advantage of the morphometric analysis is that it is rather independent of the size of the scan window.
Moreover, we discuss an example for which there is no significant excess in the total number of counts, but the source can still be detected because of the additional structural information.

Section~\ref{sec_conclusion} contains a summary of the resuls and a conclusion.
An outlook to further possible extensions is presented in Sec.~\ref{sec_outlook}.

In the appendix~\ref{sec_gamma_T}, we introduce a new test statistic, which combines different thresholds.
This allows, for example, for a better detection of diffuse radiation\footnote{Parts of this article are from the PhD thesis of one of the authors~\citep{Klatt2016}.}.

% ----------------------------------------------------------------------- %
% ----------------------------------------------------------------------- %
% ----------------------------------------------------------------------- %

\section{The statistical significance of structural deviations}
\label{sec_gamma_short_overview}

In the first paper of this series, we explained in detail the structure characterization of a counts map itself using Minkowski functionals.
Moreover, we rigorously defined the null hypothesis test that (globally) detects statistical siginficant deviations in the background noise.
The most important steps where shortly repeated in the second paper.
For a better readability, we also here repeat the definition of the test statistics.

The counts map is first turned into a black-and-white image by thresholding, that is, all pixels with a number of counts larger or equal a given threshold are set to black (otherwise white).
The Minkowski functionals of the resulting two-dimensional black-and-white image are given by the area $A$, perimeter $P$, and Euler characteristic $\chi$ of the black pixels~\citep{SchroederTurketal2010AdvMater}.
The last functional is a topological quantity.
It is given by the number of clusters minus the number of holes.

Given a measured count map, we compute for each threshold $\rho$ a triplet $(A,P,\chi)$ of Minkowski functionals.

The null hypothesis is that there are only background signals, that is, we assume that all events are randomly, independently, and homogeneously distributed over the field of view.
The first two papers demonstrated how detector effects that distort the homogeneity can be corrected for.
The number of counts follow a Poisson distribution and they are independent for different bins of equal size.
Their expectation $\lambda$ is the background intensity.

For each given threshold $\rho$, the probability distribution $\mathcal{P}(A,P,\chi)$ of the Minkowski functionals can be determined under the null hypothesis that there are only background signals, as explained in the second paper of this series.
The joint probability distribution of the Minkowski functionals sensitively characterizes the ``shape of the background noise''.
It allows to define a null hypothesis that detects statistical siginficant deviations from the background morphology.
Following~\citet{NeymanPearson1933}, we defined the \textit{compatibility} $\mathcal{C}$ of a measured triplet $(A,P,\chi)$ with the null hypothesis:
\begin{align}
  \mathcal{C}(A,P,\chi) &= \sum_{\mathcal{P}(A_i,P_i,\chi_i) \le \mathcal{P}(A,P,\chi)} \mathcal{P}(A_i,P_i,\chi_i).
  \label{eq:compatibility}
\end{align}
For convenience, we then defined the \textit{deviation strength} $\mathcal{D}$ as the logarithm of this likelihood value:
\begin{align}
  \mathcal{D}(A,P,\chi) :=& - \log_{10}\mathcal{C}(A,P,\chi);
  \label{eq:dev_strength}
\end{align}
the larger the deviation strength, the larger is the statsitical significance of the structural deviation from the background intensity, more precisely.
We reject the null hypothesis of a pure background measurement if the deviation strength is larger than $6.2$, which corresponds to the common criterion of a $\unit[5]{\sigma}$ deviation.

% ----------------------------------------------------------------------- %
% ----------------------------------------------------------------------- %
% ----------------------------------------------------------------------- %

\section{Sensitivity increase via structure characterization}
\label{sec_gamma_sensitivity}

We demonstrated that the morphometric analysis is a promising and innovative spatial data analysis in the first paper of this series, where we also discussed in an outlook its potentials and advantages.
We achieved an accurate characterization of the complex structure of the background noise in the second paper.
Here, we will show how our refined shape analysis can indeed lead to a strong increase in sensitivity.
The advanced hypothesis test based on all three Minkowski functionals can detect sources that remain undetected if only a single functional is used.

The deviation strength $\mathcal{D}(A)$ w.r.t. only the area is the following called ``simple deviation strength'',
while the deviation strength $\mathcal{D}(A,P,\chi)$ w.r.t. the complete characterization via all three Minkowski functionals is called ``joint deviation strength''.

\subsection{Examples of Minkowski sky maps}

\begin{figure}
  \centering
  \includegraphics[width=0.22\linewidth]{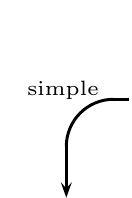}
  \subfigure[][]{%
    \includegraphics[width=0.44\linewidth]{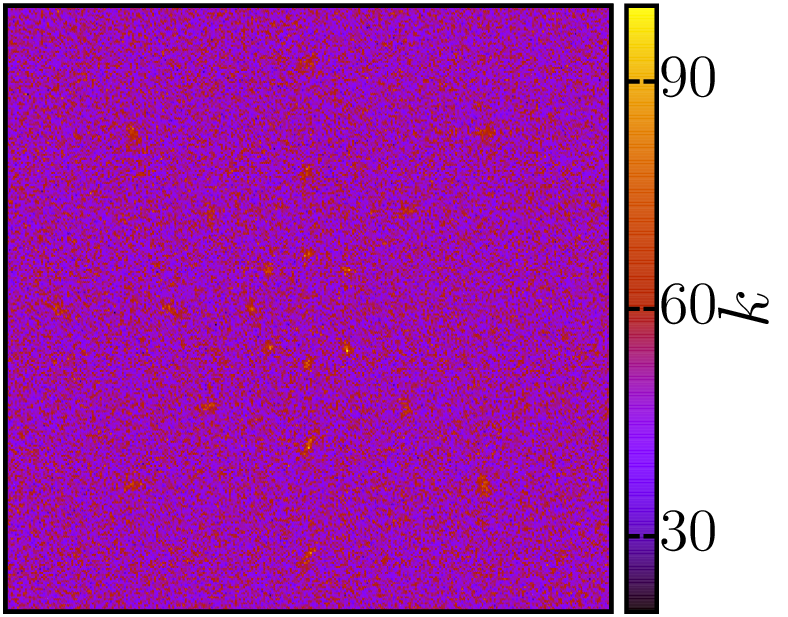}
    \label{fig_gamma_testpattern_single_count_map}
  }%
  \includegraphics[width=0.22\linewidth]{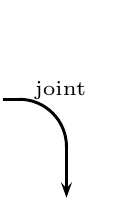}\\
  \subfigure[][]{%
    \includegraphics[width=0.44\linewidth]{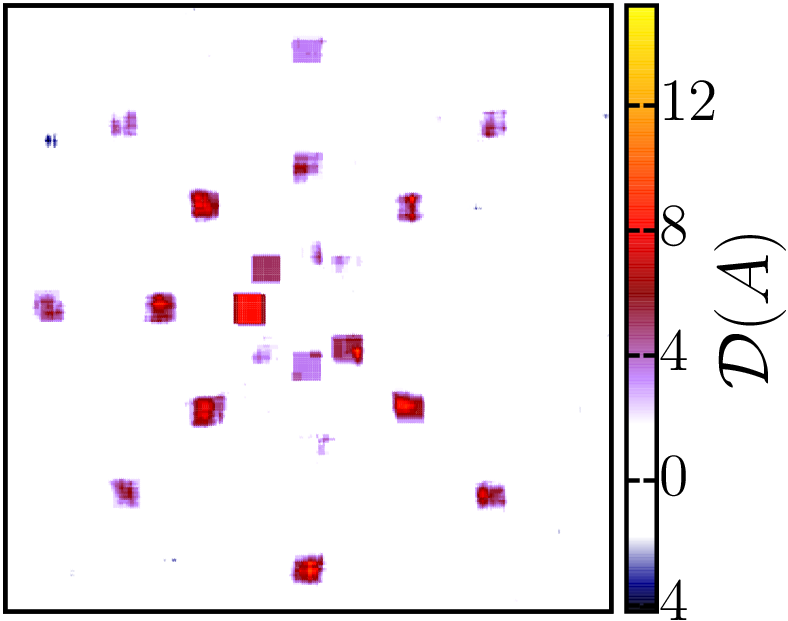}
    \label{fig_gamma_testpattern_single_simple}
  }\hspace{0.04\linewidth}%
  \subfigure[][]{%
    \includegraphics[width=0.44\linewidth]{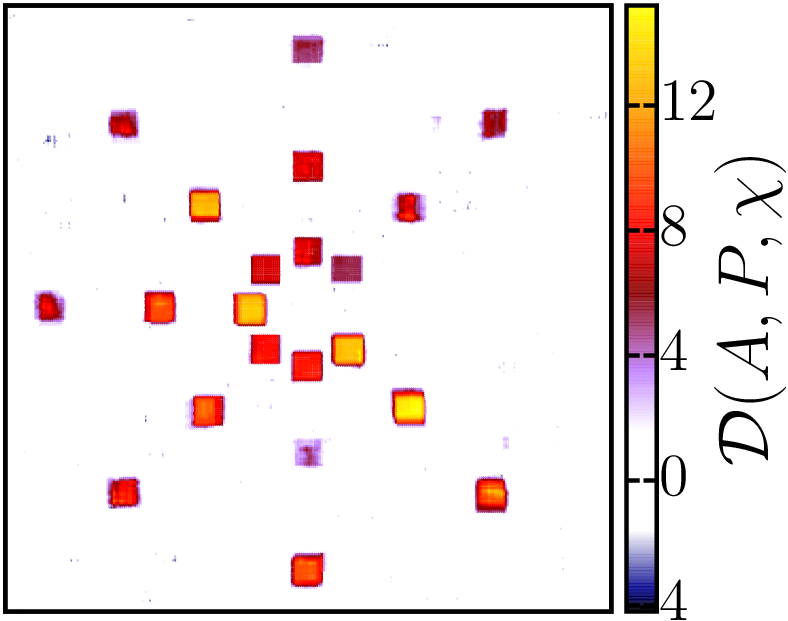}
    \label{fig_gamma_testpattern_single_joint}
  }\\
  \subfigure[][]{%
    \includegraphics[width=0.44\linewidth]{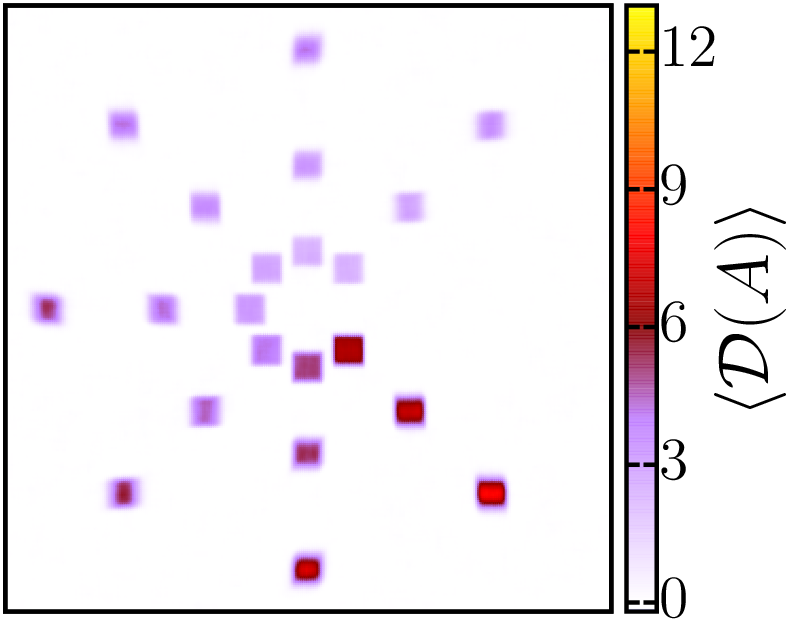}
    \label{fig_gamma_testpattern_avg_simple}
  }\hspace{0.04\linewidth}%
  \subfigure[][]{%
    \includegraphics[width=0.44\linewidth]{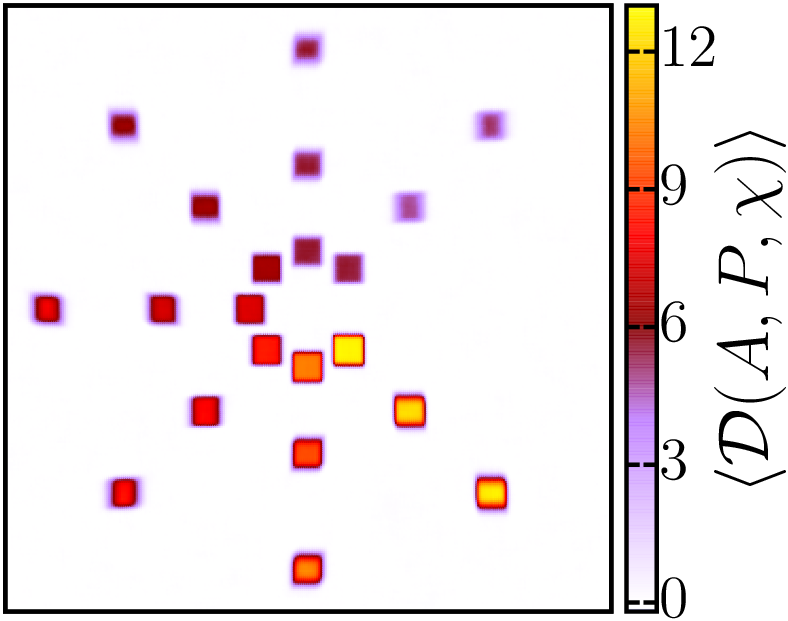}
    \label{fig_gamma_testpattern_avg_joint}
  }%
  \caption{Strong sensitivity increase via joint structure
    characterization: (a) simulated counts map including sources of
    different sizes and different integrated fluxes; the same count
    map is first analyzed by (b) a Minkowski sky map using the simple
    deviation strength, i.e., only area, then, using (c) the joint
    deviation strength, i.e., all Minkowski functionals; formerly
    undetected sources are now detected. For a more systematic
    comparison also averages of 100 Minkowski sky maps w.r.t. (d) only
    the area and (e) all three Minkowski functionals are shown.}
  \label{fig_gamma_testpattern}
\end{figure}

In order to compare the simple to the joint deviation strength, we define a test pattern including sources of different sizes and
different integrated fluxes and thus simulate count maps, see Fig.~\ref{fig_gamma_testpattern_single_count_map}.

The count map is both analyzed by a Minkowski sky map of the simple
and the joint deviation strength, see
Figs.~\ref{fig_gamma_testpattern_single_simple} and
\ref{fig_gamma_testpattern_single_joint}, respectively. Note that the
sources can be chosen much weaker than in the test pattern in the first paper of this series.
This is because the $15\times15$ sliding window uses more statistics than the small $5\times 5$ sliding
windows. Therefore, weaker sources can be detected against strong
background noise.

To demonstrate that the increased sensitivity is not a coincidence of a single fluctuation but a true grain in sensitivity due to the additional structure information,
we plot the average of 100 Minkowski sky maps based on different simulations, see
Figs.~\ref{fig_gamma_testpattern_avg_simple} and \ref{fig_gamma_testpattern_avg_joint}.
The joint deviation strengths $\mathcal{D}(A,P,\chi)$ are on average distinctly larger than the simple deviation strengths $\mathcal{D}(A,P,\chi)$ analyzing the very same data.

The simple deviation strength can be expected to be significant only for the strongest sources.
However, using all Minkowski functionals to characterize the structure of the counts maps, i.e.,
extracting more information from the same data, all sources are
detected (where the faint sources are, of course, in single simulations
sometimes detected or not depending on statistical fluctuations).
This confirms the initial idea to improve the sensitivity via an improved structure characterization.

The testpatter in Fig.~\ref{fig_gamma_testpattern} also reveals another advantage of the morphometric analysis.
Differently large sources can be detected with the same scan window size in
contrast to the standard counting method. This is discussed in more
detail in Section~\ref{sec_gamma_compare_lima}.

In an outlook based on time-consuming calculations of the DoS of a $20\times 20$, we found that the sensitivity increase gets even stronger with further increasing window sizes.
If necessary, the analysis can be extended to such window sizes.

\subsection{Systematic analysis of increase in sensitivity}
\label{sec_gamma_systematic_comparison}

The test pattern in Fig.~\ref{fig_gamma_testpattern} demonstrates that using the joint structure characterization allows for detecting
formerly undetected sources by taking additional morphometric information into account.
Of course, an increase in sensitivity can only be achieved if there actually is an additional nontrivial shape information within the scan-window, more precisely, if the shape of the source is structured on the length scale of the sliding window.

\begin{figure}
  \centering
  \includegraphics[width=\linewidth]{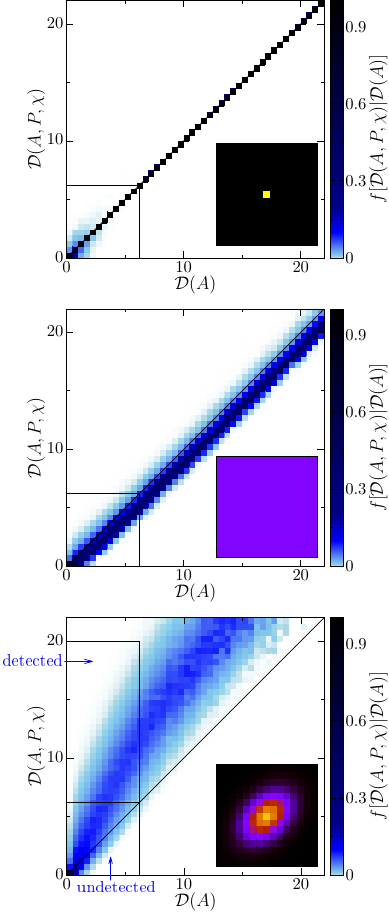}
  \caption{The sensitivity increase, comparing simple to joint deviation strengths, depends on the source shape.
    Given $\mathcal{D}(A)$, the color code shows the frequency $f$ of $\mathcal{D}(A,P,\chi)$ for the same counts map.
    We analyze (top) a true point source that is smaller than a pixel, (center) a uniform offset compared to the
    background intensity, and (bottom) an extended source with a
    nontrivial shape within the scan window, i.e., a strong
    gradient in the intensity. The insets show the intensity profiles of the sources.}
  \label{fig_gamma_sensitivity}
\end{figure}

If any morphometric approach is to analyze a completely uniform offset in the background intensity, i.e., a Poisson random field with a different intensity $\lambda'>\lambda$,
the result must be less significant than for a simple method based only the total number of counts.
This is simply because the additional structural information is in perfect accordance with the background model and the only difference is a different total number of counts.

In Fig.~\ref{fig_gamma_testpattern}, the simple and joint deviation strengths are systematically compared to each other for differently shaped sources:
\begin{enumerate}
  \item a true point source which is smaller than a pixel,
  \item a uniform offset in the background intensity,
  \item and a Gaussian shaped source.
\end{enumerate}
Between 0.75 and 7.5 million counts maps ($15\times 15$) are simulated using the same intensity profile but different integrated fluxes.
For each count map, both the simple and the joint deviation
strength are determined. Given a simple deviation strength
$\mathcal{D}(A)$, the conditional frequency
$f[\mathcal{D}(A,P,\chi)|\mathcal{D}(A)]$ of the joint deviation
strength $\mathcal{D}(A,P,\chi)$ is determined, i.e., for all cells
with a simple deviation strength $\mathcal{D}(A)$ the empirical
probability density function of the joint deviation strength
$\mathcal{D}(A,P,\chi)$ is computed. The result is plotted using a
color scale in Fig.~\ref{fig_gamma_testpattern}.
The black diagonal line indicates what would be equal values of simple and joint deviation strength.
The vertical and horizontal black lines depict the null hypothesis criterion for the simple or joint deviation strength, respectively.

Not even the strongest true point source in these simulations would
have been detected by a simple counting method because the source
signals are suppressed by the large additional background. In contrast
to this, even the simple deviation strength uses additional
information the dependence on the threshold $\rho$ and can thus, e.g.,
detect a single pixel with an exceptional high number of counts
because of a unlikely black pixel at very high thresholds. This
advantage in being more independent on the system size is discussed in
more detail below in Section~\ref{sec_gamma_compare_lima}.

However, comparing $\mathcal{D}(A)$ to $\mathcal{D}(A,P,\chi)$, there
is no additional information in the perimeter $P$ or Euler
characteristic $\chi$: If there is only a single black pixel at high
thresholds, the only possible values for $P$ and $\chi$ are 4 or 1,
respectively. Therefore, are the simple and joint deviation strengths
are exactly identical.

For the uniform source, the additional information ($P, \chi$) must,
as stated above, lead to a decrease of the deviation
strength. Interestingly, this decrease turns out to be rather small
even in the extreme case of a constant offset. The decrease of the
deviation strength could even be an advantage in that the joint
deviation strength is slightly less sensitive to errors in the
estimation of the background intensity and a source would still be
detected because of the strong deviation in the area $A$.

For the structured source, there is a tremendous increase in
sensitivity for the joint deviation strength compared to the simple
one.
For all counts maps for which the corresponding values of the
deviation strengths are within the dashed box, the source is not
detected if
only the area characterizes the structure, but it is detected by the
joint deviation strength, i.e., if all Minkowski functionals
characterize the shape of the counts map. For the same counts map for
which the simple deviation strength based only on the area is below 5,
i.e., the compatibility is more than $10^{-5}$, the joint deviation
strength reaches values nearly 20, i.e., with compatibilities less
than $10^{-19}$. In other words, if the structure is characterized not
only by the area but by all Minkowski functionals, the compatibility
with the background structure can drop by 14 orders of
magnitude. There is no significant excess in the total number of
counts but in the structure of the counts map. Only by taking this
morphometric information into account, a formerly undetected source
can now be detected using the same data.

\begin{figure}
  \centering
  \includegraphics[width=\linewidth]{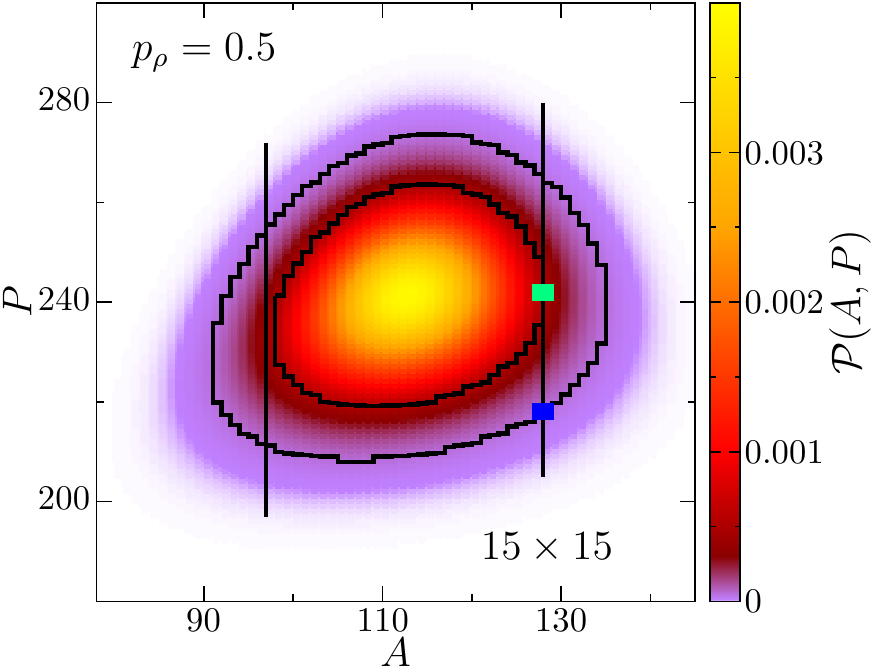}
  \caption{Probability distribution $\mathcal{P}$ of area $A$ and
    perimeter $P$ of a $15\times15$ b/w image with probability
    $p_{\rho}=0.5$ for a pixel being black. }
  \label{fig_gamma_explanation_shape_dependence}
\end{figure}

A more formal explanation of this intuitive understanding can be given
with the aid of
Fig.~\ref{fig_gamma_explanation_shape_dependence}. Given a measured
area, e.g., $A=128$, the compatibility $\mathcal{C}(A)$ w.r.t. only
the area is the sum over all probabilities left of the left dotted
line and right of the right dotted line, i.e., all macrostates with an
area less likely than the given area $A=128$.

In the example of a uniform offset in the background intensity, the
structure, quantified by the perimeter\footnote{Because of the white
    boundary conditions the perimeter can only take on even values; for an odd value of $P$ the probability is zero. For an easier visualization the bin length in $P$ is two.}
 $P$, is in agreement with the
background structure and the perimeter most likely takes on a
``typical'' value, i.e., a likely perimeter for a given area $A$,
e.g., the green square represents $P=242$. The compatibility
$\mathcal{C}(A,P)$ is then the sum over all probabilities outside the
inner contour line, which results in $\mathcal{C}(A,P) >
\mathcal{C}(A)$.
However, in the example of a structured source the perimeter $P$ might
take on an unlikely value for the perimeter, e.g., the blue square
represents $P=218$. The compatibility $\mathcal{C}(A,P)$ is now the
sum over all probabilities outside the outer contour line and thus,
$\mathcal{C}(A,P) < \mathcal{C}(A)$. A structural deviation from the
background structure leads to a more significant result of the
morphometric analysis compared to simple counting methods. For two
different b/w images with the same compatibility with the background
w.r.t. the area, the additional information of the perimeter specifies
whether the b/w image is indeed compatible to the background structure
or not.

% ----------------------------------------------------------------------- %
% ----------------------------------------------------------------------- %
% ----------------------------------------------------------------------- %

\section{Comparison to a standard counting method}
\label{sec_gamma_compare_lima}

The standard null hypothesis test in gamma-ray astronomy was
introduced in \citet{LiMa1983}: it compares the
number of signals $N_{on}$ in the so-called ``on-region'', i.e., in
the vicinity of an expected source, to the number of background
signals detected in an $N_{\mathit{off}}$ ``off-region'', i.e., a
region in the sky without sources. The method simply counts the number
of photons. Given an exposure ratio $\alpha$, the significance
$\sigma$ is
\begin{align}
  \sigma = \sqrt{ \ln\left\{\left[ \frac{1+\alpha}{\alpha} \left(\frac{N_{\mathit{on}}}{N_{\mathit{on}}+N_{\mathit{off}}} \right) \right]^{2 N_{\mathit{on}}}
                            \left[ (1+\alpha) \left(\frac{N_{\mathit{off}}}{N_{\mathit{on}}+N_{\mathit{off}}} \right) \right]^{2N_{\mathit{off}}}\right\}
                }.
\end{align}
The significance can be expressed in terms of a deviation strength~\citep{GoeringKlattEtAl2013}:
\begin{align}
    \mathcal{D}(\sigma)  &= -\log_{10} \left( \mathrm{erfc} \left( \frac{\sigma}{\sqrt{2}} \right) \right)
\label{eq_gamma_D_of_sigma}
\end{align}
where $\mathrm{erfc}(x) = \frac{2}{\sqrt{\pi}} \int_x^\infty{\exp(-t^2) \; \mathrm{d}t}$ is the error function.
This standard counting method allows for a simple and fast null
hypothesis test. However, the analysis only uses the total number of
counts in the observation window.

Our morphometric analysis follows a completely new ansatz based on
structure characterization, which can not only detect an excess of
counts but is able to detect any structural deviation from the
background noise.
It can detect structural deviations in count maps where the number
of expected counts is in perfect agreement with the background intensity~\citep{Klatt2016}.

\subsection{Dependence on experimental details}

Therefore, there is no direct and straightforward comparison that one
of the methods is always more sensitive than the other. A comparison of
the advantages and different possibilities of the two methods is
complicated and depends on the experimental details and the source
shape.
The standard counting method is more likely to detect sources if there
is no interesting structure to be quantified within the scan window.
The additional structural information is in agreement with the
background noise, as discussed in
Section~\ref{sec_gamma_systematic_comparison}. The morphometric
analysis based on all Minkowski functionals is, in this case, less
likely to detect a source compared to an analysis that only takes the
total number of counts into account.
However, if there is a distinct structural difference from the
background noise, the morphometric analysis has the advantage of being
able to use this additional information.

Whether or not there is an increase in sensitivity compared to the
counting method by Li and Ma also strongly depends on the
experimental details, e.g., the bin size, as shortly discussed in
\citet{Goering2012}. In contrast to the total number of counts,
our morphometric analysis strongly depends on the choice of the bin
size. If the bins are too large, interesting source structure might be
hidden because it is contained in a single bin. However, if the bin
size is too small, the sky map does not only get very noisy but we can
also loose structural information. In the extreme case that in the
black and white image all black pixels are separated from each other
by white pixels, the translation invariant Minkowski functionals can
no longer distinguish different configurations, but only the total
number of black bins. Therefore, the bin size needs to be chosen
reasonably taking the size of the scan window, the point spread
function of the telescope, the quality of the data, and the source
shape into account, see \citet{Goering2012}.

% Accuracy of lambda-estimate
As mentioned above, the morphometric analysis can detect sources even
if there is no excess in the signals compared to the background
intensity, see
Section~\ref{sec_gamma_systematic_comparison}. Therefore, we expect the
morphometric analysis to be robust against overestimates of the
background intensity $\lambda$. However, a more thorough analysis of
such effects for real data is beyond the scope of this article.
Ideally, it would need extensive simulations to determine the
empirical cumulative distribution function with a priori unknown
$\lambda$, but instead using an efficient estimation of the background
intensity, as described in \citet{Goering2012}.
A main advantage of the morphometric analysis compared to the counting
method could eventually be that it avoids observations of off-region
because a less precise estimate of the background intensity is
sufficient.

The method by Li and Ma compares the number of counts in the source
region and in regions with only background signals. Obviously, it
strongly depends on how accurate the estimate of the background
intensity is.
Here, we compare the morphometric analysis to the significance of the Li
and Ma test for the extreme and most sensitive case of an infinitely
long observation of the off-region\footnote{An infinite observation
  time corresponds to using the exact background intensity.}
$N_{\mathit{off}}\rightarrow \infty$ while
$\alpha=\lambda_{tot}/N_{\mathit{off}}\rightarrow 0$. In this limit,
\begin{align}
  \sigma = \sqrt{2}\left\lbrace N_{\mathit{on}} \ln\left[
      \frac{N_{\mathit{on}}}{\lambda_{tot}} \right] + \lambda_{tot} - N_{\mathit{on}}
  \right\rbrace^{1/2}\, .
  \label{eq:08_lima}
\end{align}

For a final comparison of both methods, their dependencies on these
experimental details must be accurately studied, which is beyond the
scope of this article. Here, we discuss the advantages of the refined
morphometric analysis and compare it to the counting method in two
examples, where the morphometric analysis detects sources in contrast
to the standard method by Li and Ma. These examples show the potential
of the morphometric ansatz. However, the choice of the optimal method
strongly depends on the details of the data which is to be analyzed,
as well as, on the information or features that are to be extracted.

\subsection{Scan window size dependence and low statistics}

% window size dependence
Besides the above mentioned robustness against overestimated
background intensities or the ability to even detect inhomogeneities
with no excess in the total number of counts, another important
advantage of the morphometric analysis is that it depends much less on
the choice of the size of the scan window. 
Even sources with extensions much smaller than the scan window size
are detected in the morphometric analysis, although there is no
significant change in the total number of counts.
 
Figure~\ref{fig_gamma_sensitivity} shows how even a point source,
i.e., a single pixel with increased intensity, can sensitively be
detected, because at high threshold even a single black pixel is very
unlikely, which is independent of counts in other pixels possibly in
agreement with the null hypothesis.
The morphometric analysis can detect sources of very different
extensions with the same scan window size.

In contrast to this, the standard counting method cannot detect a
source if there are at the same time too many pixels with only
background signals in the same scan window. If there are many pixels
within the scan window which contain only background signals, the
source signals can be suppressed. The small excess in the total number
of counts is no longer significant. This effect can be reduced if the
size of the scan window is adjusted to the extension of the
source. However, such an adaption can possibly lead to a biased choice
of the parameter or an unknown trial factor if the analysis is
repeated with different window size.

\begin{figure}[t]
  \centering
  \hfill%
  \subfigure[][]{%
    \includegraphics[width=0.48\linewidth]{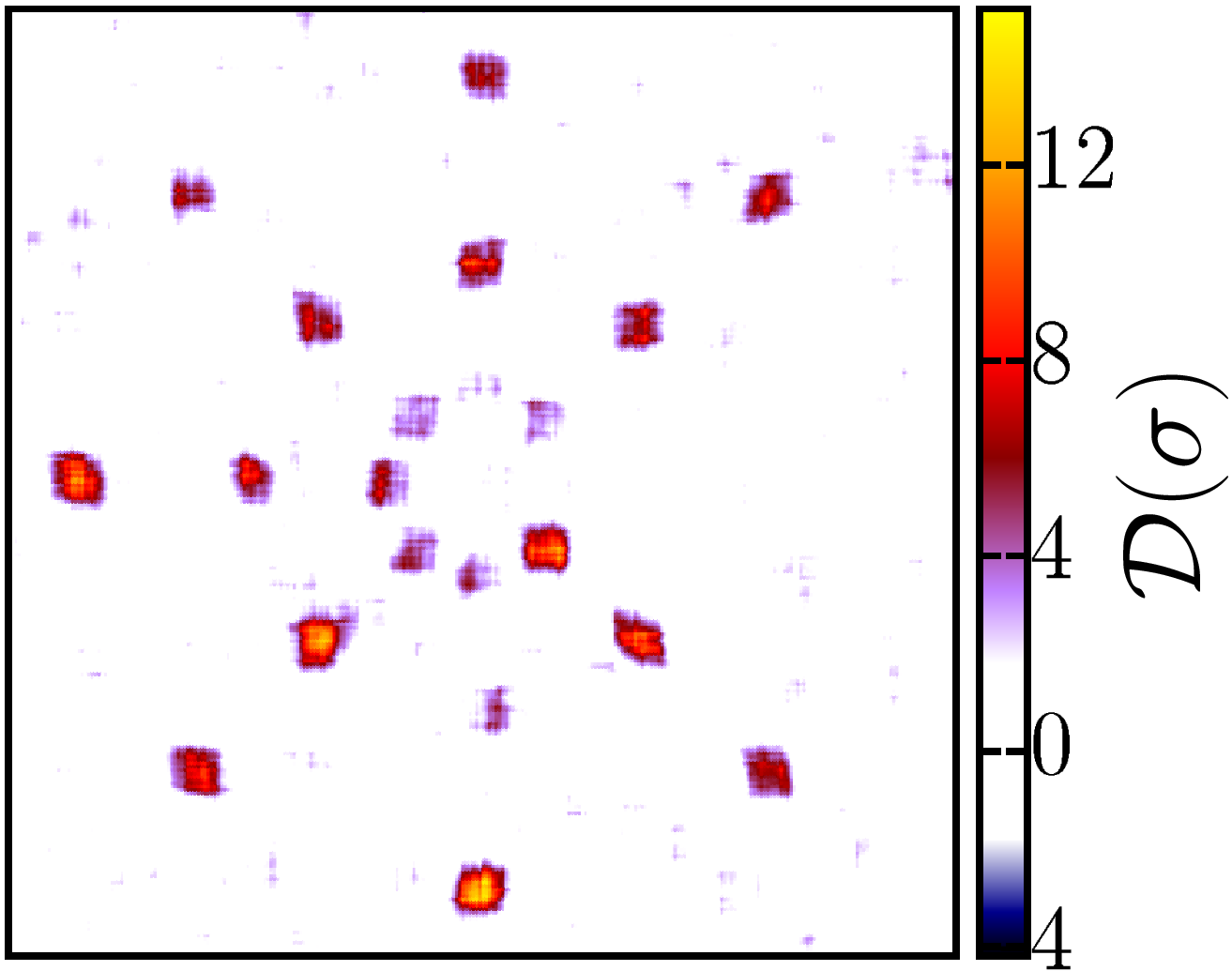}
  }%
  \hfill%
  \subfigure[][]{%
    \includegraphics[width=0.48\linewidth]{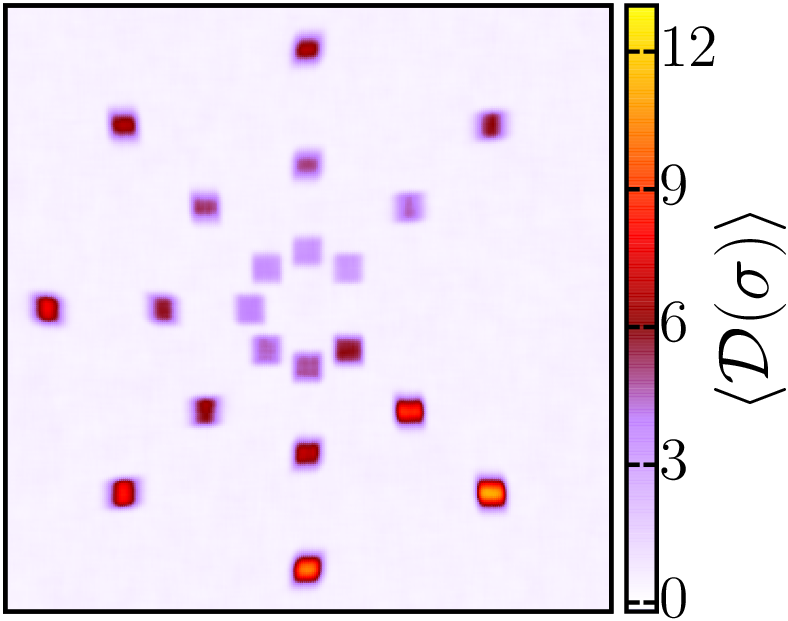}
  }%
  \hfill%
  \caption{For the test pattern in Fig.~\ref{fig_gamma_testpattern},
    the excess in the total number of counts is determined; the
    deviation strength in the total number of counts is derived from
    the significance of the excess according to
    Eqs.~\eqref{eq_gamma_D_of_sigma} and \eqref{eq:08_lima}: (a) a
    single sky map is shown like in
    Figs.~\ref{fig_gamma_testpattern_single_simple} or
    \ref{fig_gamma_testpattern_single_joint}, (b) an average of 100
    Minkowski sky maps like in
    Figs.~\ref{fig_gamma_testpattern_avg_simple} or
    \ref{fig_gamma_testpattern_avg_joint}; for a direct comparison,
    the same color scales are used. For the chosen $15\times15$ scan
    window, the outer sources are detected with a similar significance
    as in Fig.~\ref{fig_gamma_testpattern}. However, in order to
    detect the inner sources, the scan window size must be adjusted.}
  \label{fig_gamma_testpattern_sigma}
\end{figure}

In Fig.~\ref{fig_gamma_testpattern_sigma}, the test pattern from
Fig.~\ref{fig_gamma_testpattern} with differently large sources is
analyzed using the same scan window size as the morphometric analysis,
see
Figs.~\ref{fig_gamma_testpattern_single_simple}--\ref{fig_gamma_testpattern_avg_joint}.
The large outer sources are of the same size as the scan window; they
are detected with a similar significance as by the morphometric
analysis. However, the smaller inner sources are not statistically
significantly detected, because there are too many background signals
in the same scan window. If the scan window size is adjusted, these
sources can be detected highly significantly. However, the problem of
a possibly biased choice of parameters, as well as, an unknown
trial factor remains.

The most important advantage of the morphometric analysis is, of
course, that it incorporates additional structural
information. Especially, if only low statistics are available, an
increase in the sensitivity by quantifying the structure of the counts
map is probably most needed. For example, a slight excess in the total
number of counts might not be significant because of the strong
fluctuations of a Poisson distribution relative to the small mean
value. Interestingly, especially in such a case the significance in
the structural deviation is relatively strong compared to the
significance in the number of counts. For example, for a Poisson
random field with a low intensity the clustering of a given number of
black pixels is equally likely or unlikely as in a field with high
intensity. The excess in the number of counts might not be
significant, but the improbable arrangement of the black pixels can
lead to the detection of the source.
So, the advantage of additional structural information should be most
effective when it is most needed.

\begin{figure}[t]
  \centering
  \includegraphics[width=\linewidth]{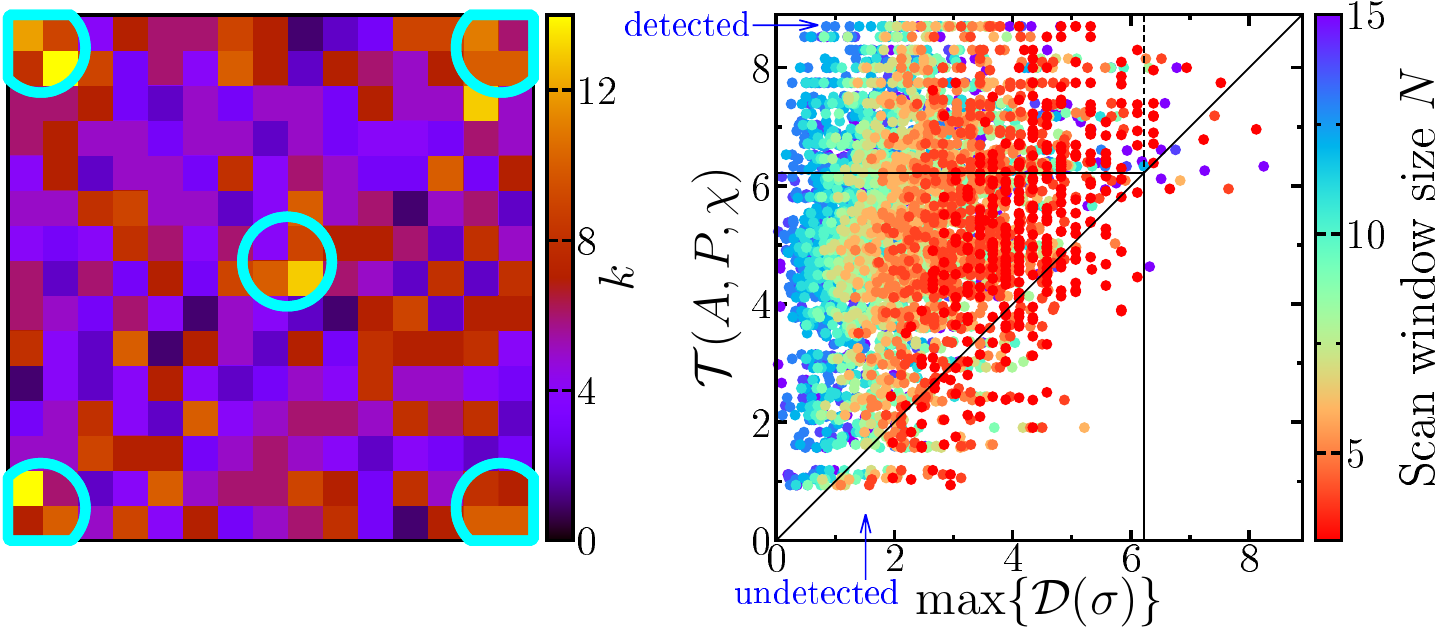}
  \caption{A source like a dotted pattern where only low statistics
    are available: (a) the source consists of several pointlike
    sources (marked by circles) hardly visible by eye in the counts
    map; (b) comparison of the deviation strength
    $\mathcal{T}(A,P,\chi)$ of the morphometric analysis to
    $\mathcal{D}(\sigma)$ of the standard counting method from
    Eqs.~\eqref{eq_gamma_D_of_sigma} and Eq.~\eqref{eq:08_lima}. 400
    samples are simulated. For each sample, the morphometric analysis
    is applied to the whole $15\times15$ counts map. For the same
    sample, it is compared to $\mathcal{D}(\sigma)$, which is
    evaluated for all scan window sizes between $15\times15$ (blue
    points) and $3\times 3$ (red points) where the maximum of all iterations
    over the counts map is used. Although the trial factors for
    $\max\{\mathcal{D}(\sigma)\}$ are ignored, the morphometric
    analysis is for the vast majority of samples more sensitive
    $\mathcal{T}(A,P,\chi)>\max\{\mathcal{D}(\sigma)\}$. Although
    there might be no significant excess in the total number of
    counts, the source can be detected by taking more information out
    of the same data.}
  \label{fig_gamma_compare_D_to_T_speckle_systematic}
\end{figure}

Figure~\ref{fig_gamma_compare_D_to_T_speckle_systematic} shows such an
example of a weak and hardly detectable source where only low
statistics are available. Although hardly visible by eye in the count
map, there are strong intensity gradients. The source resembles a
dotted pattern; it consists of several nearly pointlike sources that
are marked in the count map on the left-hand side. Because of the
strong intensity gradients and thus the significant structural
deviation from the background noise, the morphometric analysis can
take advantage of the additional geometrical information.

Because of the low statistics, there are strong statistical fluctuations in the single sky maps.
We therefore do here not use the maximal deviation strength over all thresholds, but an improved test statistic that combines the deviations strength of different thresholds, as explained in the Appendix~\ref{sec_gamma_T}.
For a more systematic analysis, I
simulate 400 samples and compare the deviation strength
$\mathcal{T}(A,P,\chi)$ of the morphometric analysis to
$\mathcal{D}(\sigma)$ of the standard counting method from
Eqs.~\eqref{eq_gamma_D_of_sigma} and \eqref{eq:08_lima}, see the
right-hand side of
Fig.~\ref{fig_gamma_compare_D_to_T_speckle_systematic}.

The morphometric analysis analyzes the whole $15\times15$ count
map. However, it is not only compared to $\mathcal{D}(\sigma)$ of the
standard counting method using the same scan window size (blue
points), rather for all scan windows down to very small sizes $3\times
3$ (red points). Thus, the effect of windows size dependence in the
standard counting method can be taken into account. For scan windows
smaller than the total size of the counts map, we iterate the scan
window over the count map and compare the maximum of all deviation
strengths $\mathcal{D}(\sigma)$ to $\mathcal{T}(A,P,\chi)$.

Although the thus necessary trial factors for
$\max\{\mathcal{D}(\sigma)\}$ are ignored, which in some cases would
reduce the deviation strength by more the $1.4$, the morphometric
analysis is for the vast majority of samples more sensitive than the
counting method
$\mathcal{T}(A,P,\chi)>\max\{\mathcal{D}(\sigma)\}$. While there are
only about 5 out of 400 samples with $\max\{\mathcal{D}(\sigma)\} >
6.22$, there are many samples where the source is not detected by
$\max\{\mathcal{D}(\sigma)\}$ but by $\mathcal{T}(A,P,\chi)$.  In
other words, although there might be no significant excess in the
total number of counts, the source can still be detected by taking
more information out of the same data.
In general, a comparison of the morphometric analysis and the
standard method by Li and Ma depends on both the experimental details
and the structure of the source. For this example, the morphometric
analysis is more sensitive.

% ----------------------------------------------------------------------- %

\section{Conclusion}
\label{sec_conclusion}

The morphometric analysis allows to detect gamma-ray sources via structural deviations from the background noise without any assumptions about potential sources.

Comparing the simple to the joint structure characterization, we can demonstrate a significant increase in sensitivity due to the additional shape information that is extracted from the data, see Fig.~\ref{fig_gamma_testpattern}.
For the same counts map for which the simple deviation strength based only on the area is below 5, i.e., the compatibility is more than $10^{-5}$, the joint deviation strength can reach values of nearly 20, i.e.,
compatibilities less than $10^{-19}$, see Fig.~\ref{fig_gamma_sensitivity}.
The compatibility with the background structure drops by 14 orders of magnitude and formerly undetected sources can be detected simply by applying a refined morphometric analysis.
Of course, the increase in sensitivity depends on the shape of the source, see Fig.~\ref{fig_gamma_explanation_shape_dependence}.
 
A comparison of the morphometric analysis to the standard null
hypothesis test by Li and Ma in gamma-ray astronomy, see
Eqs.~\eqref{eq_gamma_D_of_sigma} and \eqref{eq:08_lima}, depends both
on the shape of the source and on the experimental details, like the
binning or the accuracy of the estimate of the background
intensity. The morphometric analysis follows a very different
ansatz. Besides the advantage of including additional structural
information, it depends less on the size of the scan window and can
detect both rather extended and pointlike sources using the same scan
window size, compare Figs.~\ref{fig_gamma_testpattern} and
\ref{fig_gamma_testpattern_sigma}. Moreover, the advantage of
additional structural information should be most effective for short
observation times and low
statistics. Figure~\ref{fig_gamma_compare_D_to_T_speckle_systematic}
shows an example for which there is no significant excess in the total
number of counts, but the source can still be detected because of the
additional structural information.

In summary, the main advantages of the morphometric analysis are:
\begin{itemize}
  \item a shape analysis without prior knowledge about potential sources,
  \item a sensitivity gain via structure information especially for low statistics,
  \item its relative independence of the scan window size, and
  \item its detection of statistically significant inhomogeneities in the counts map even if the expected total number of counts is in perfect agreement with the background intensity.
\end{itemize}
Moreover, we expect for the reasons discussed above that it is robust against errors in the estimation of the background intensity.

The simulation study in this article, demonstrates how additional information extracted from the same data can allow the detection of formerly undetected sources.
The next step is to apply these improved techniques to real data from experiments; for first examples of applications to H.E.S.S. sky maps, see \citet{Goering2008, Goering2012, Klatt2016}.

% ----------------------------------------------------------------------- %

\section{Outlook}
\label{sec_outlook}

The morphometric analysis is here shown to be an innovative and efficient spatial data analysis.
Of course, there are even further possibilities to extend the analysis.
For example, the method can naturally be extended to any spatial dimension if the structure of the $\mathrm{D}$ dimensional b/w image is characterized by the $(\mathrm{D}+1)$ Minkowski functionals.

\subsection{Other shape descriptors}
\label{sec_gamma_other}

The above-defined analysis is very general, and indeed any useful shape descriptor can be used.
The Minkowski functionals are versatile tools and can comprehensively quantify the structure of quite different random fields.
However, if for a certain system another index is better physically motivate, more important or interesting, it can replace the Minkowski functionals, but the basic idea remains unchanged.
Only the probability distribution, i.e., the DoS has to be determined following the procedure from the second paper in this series.

The Minkowski functionals already incorporate all additive and conditional continuous scalar geometrical information, and we have already also introduced the Minkowski tensors to the morphometric analysis in \citet{Klatt2010}.
Other shape descriptors like the convexity number~\citep{Stoyan1987} or Betti numbers~\citep{Robins2002} are directly applicable to the method describe above, but the calculations will be more time consuming.

Even functions can be used as shape characteristics.
A first example could be the cluster function that is the probability of finding two points at a given distance in the same cluster.
Only one additional step is needed: the correlation function must be mapped to a scalar.
A good choice would be the integral over the absolute value of the difference to the mean value (i.e., expected) correlation function of a Poisson random field; this is well defined if there are no long-range correlations, because then all correlation functions converge sufficiently fast to the same constant and thus, the difference to zero.
An even more interesting example could be correlation functions of Minkowski functionals~\citep{MeckeBuchertWagner1994, KlattTorquato2014}.

\subsection{Further random fields}
\label{sec_gamma_corr_betw_pixels}

We have developed the morphometric analysis for analyzing counts map in gamma-ray astronomy, where the counts in different pixels are uncorrelated because they result from different events (showers) clearly separated in time.
We here show how the morphometric analysis distinguishes homogeneous from inhomogeneous Poisson random fields.
However, the morphometric analysis allows for a much more general analysis.
It can detect other deviations from a Poisson assumption, e.g., correlations between the counts in different pixels; see Section~\ref{sec_gamma_corr_betw_pixels}.
There might be no deviation in the number of counts (globally or even locally), but a strong deviation in the structure quantified by the Minkowski functionals.

Although we have developed the morphometric analysis for analyzing counts map with uncorrelated pixels, its use might even be more efficient for other applications with correlations between the counts in different pixels,
for example, in detectors where an event is simultaneously triggered in neighboring pixels.

Note also, that the concept of the morphometric analysis can immediately be extended to other random fields, e.g., the Boolean models or the Gaussian random field.
The Minkowski functionals are, as mentioned above, already used to search for statistical significant deviations from a Gaussian random field in the cosmic microwave background; for example, see~\citep{Schmalzing1999,Gay2012,Ducout2013}.
Of course, the probability distributions of the Minkowski functionals need to be determined and probably only numerical estimates are possible.

\subsection{Extensions of the test statistic}

In the appendix, we introduce a new test statistic combining different thresholds, see Eq.~\eqref{eq_gamma_def_T}.
Instead of the maximum of the deviation strength, we use the sum of the deviation strengths over all thresholds, see Eq.~\eqref{eq_gamma_def_S}.
We determine the empirical complementary cumulative distribution function, see Fig.~\ref{fig_gamma_epdf}.
The combination of the structural information at different thresholds improves especially the detection of diffuse radiation and extended sources, see Fig.~\ref{fig_gamma_compare_D_to_T}.

\begin{acknowledgements}
  We thank Christian Stegmann and Daniel G\"oring for valuable discussions, suggestions, and advice.
  We thank the German Research Foundation (DFG) for the Grant No. ME1361/11 awarded as part of the DFG-Forschergruppe FOR 1548 ``Geometry and Physics of Spatial Random Systems.''
\end{acknowledgements}

\appendix

% ----------------------------------------------------------------------- %
% ----------------------------------------------------------------------- %
% ----------------------------------------------------------------------- %

\section{Combining different thresholds}
\label{sec_gamma_T}

\begin{figure}[t]
  \centering
  \subfigure[][]{%
    \includegraphics[height=4.1cm]{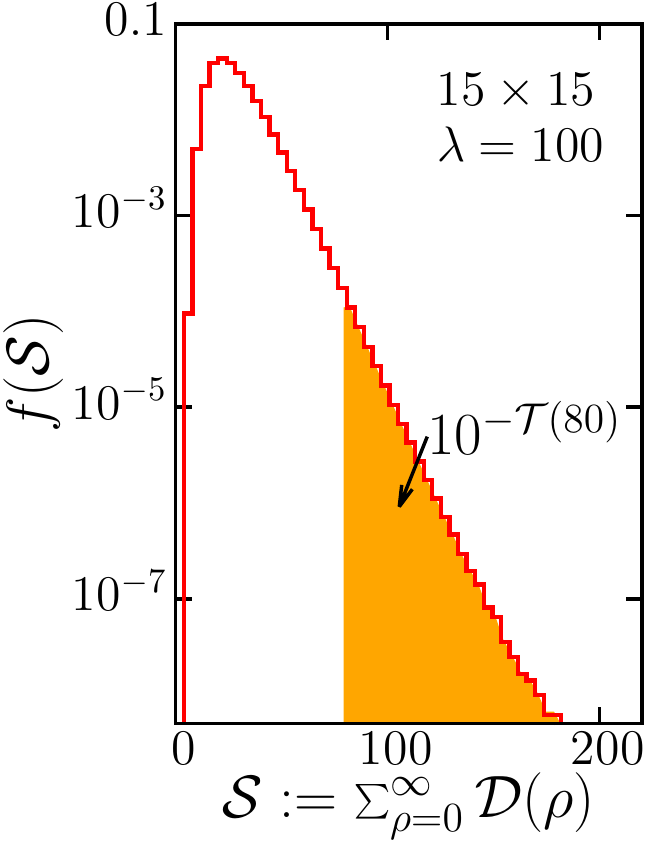}
  }%
  \hspace{-2pt}
  \subfigure[][]{%
    \includegraphics[height=4.1cm]{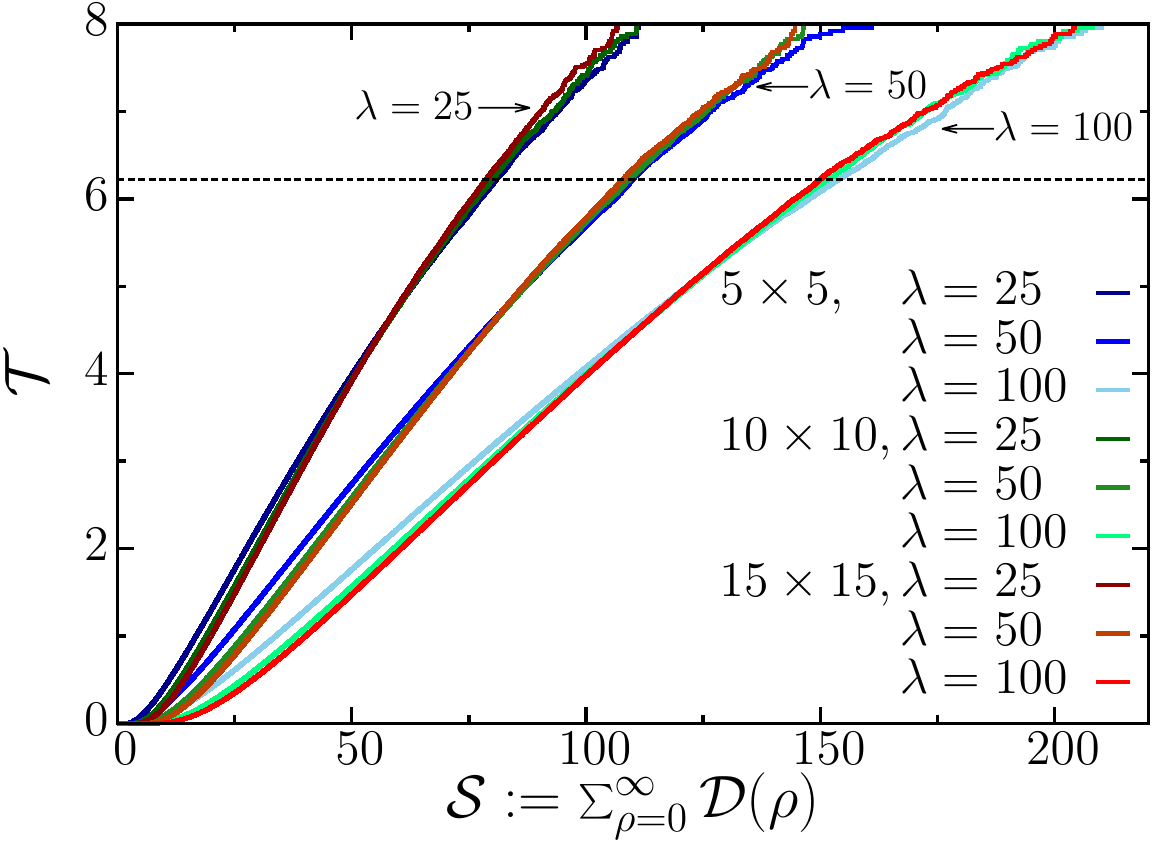}
  }%
  \caption{Sum of deviation strengths $\mathcal{D}(\rho)$ over all
    thresholds $\rho$: (a) empirical probability density function
    $f(\mathcal{S})$ and (b) the new test statistic $\mathcal{T}$,
    which is the negative decadic logarithm of the empirical
    complementary cumulative distribution function for different
    system sizes and background intensities $\lambda$; the dashed line
    indicates the hypothesis criterion, which is adjusted to the
    common $5\sigma$ criterion.}
  \label{fig_gamma_epdf}
\end{figure}

So far, only the deviation strength of a single threshold is directly
used for the null hypothesis test. The deviation strengths at other
thresholds are used only indirectly by the fact that they are smaller
than the maximum. However, the deviation strength as a function of the thresholds
contains a lot of information, e.g., even to some degree the extension
of the source~\cite[][Fig.~3]{GoeringKlattEtAl2013}. Taking this information into account could yield a
profound additional insight in the spatial data. However, it is
currently out of reach to determine the probability distribution of
the deviation strength as a function of the threshold.

Nevertheless, the most important information is whether the maximal
deviation strength is only a fluctuation at a single threshold or
whether there are strong structural deviations over a large range of
thresholds~\cite[][Fig.~3]{GoeringKlattEtAl2013}. This can
be quantified by replacing the maximum of the deviation strength by
the sum of all thresholds\footnote{The infinity norm $l_{\infty}$ is
  replaced by the 1-norm $l_1$.},
\begin{align}
  \mathcal{S}:=\sum_{\rho=0}^{\infty}\mathcal{D}(\rho)\, ,
  \label{eq_gamma_def_S}
\end{align}
which is well defined, i.e., $\mathcal{S} < \infty$, because for every
counts map there is a maximum count $k_{\mathrm{max}}$; thus, for
$\rho > k_{\mathrm{max}}$ the b/w image is completely white. For a
large enough threshold $\rho_{l} \gg 1$, this becomes the most likely
configuration, because if $p\rightarrow 0$,
$\mathcal{P}(A=0)=(1-p)^{N^2}\rightarrow 1$. For $\rho > \rho_{l}$,
the compatibility is one and the deviation strength zero; the series,
defined in Eq.~\eqref{eq_gamma_def_S}, is actually a finite sum.

The sum $\mathcal{S}$ is a new test statistic instead of the maximum
deviation strength $\mathcal{D}$ before. The distribution of this new
test statistic can impossibly be calculated analytically, but efficient
and tight approximations might be achievable, although out of the scope of
this article. Here, the cumulative distribution is determined
numerically, and the sensitivity gain is shown for simulated data.

\subsection{Empirical cumulative distributions}
\label{sec_gamma_ecdf}

\begin{figure}[t]
  \centering
  \includegraphics[width=\linewidth]{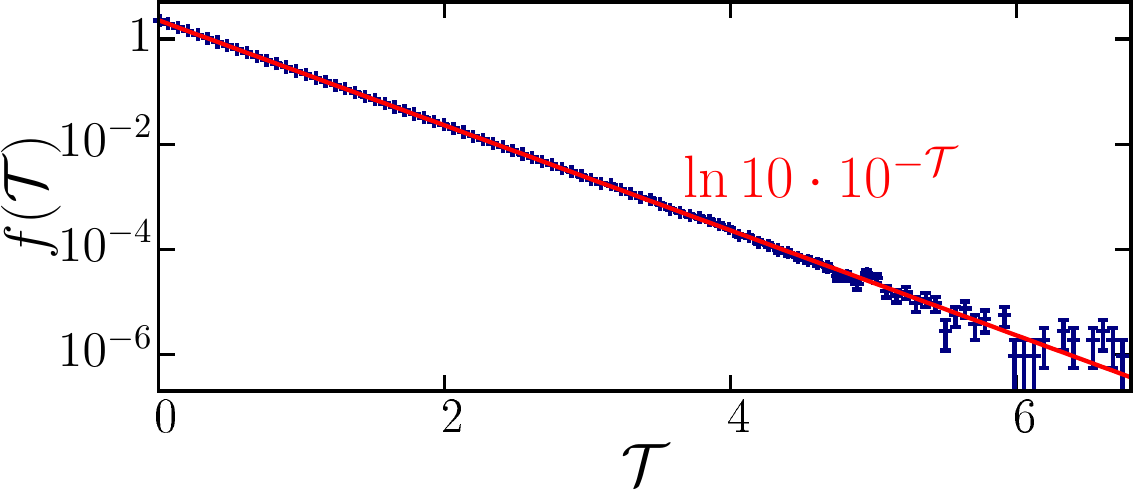}
  \caption{Empirical probability density function of the test statistic
    $\mathcal{T}$ from Eq.~\eqref{eq_gamma_def_T} for a $15\times 15$
    count map with a background intensity $\lambda=50$; it is in very
    good agreement with the target probability distribution from
    Eq.~\eqref{eq_gamma_distribution_of_T}.}
  \label{fig_gamma_distribution_of_T}
\end{figure}

We simulate $10^9$ counts maps and for each calculate
$\mathcal{S}:=\sum_{\rho=0}^{\infty}\mathcal{D}(\rho)$, from which I
derive the Empirical Probability Density Function (EPDF) for different
window sizes and background intensities. Figure~\ref{fig_gamma_epdf}
shows the EPDF $f(\mathcal{S})$ for a $15\times 15$ Poisson counts map
with intensity $\lambda=100$.

The new test statistic is defined as the empirical complementary
cumulative distribution function (ECCDF), i.e., given a measured sum
of deviation strength $\mathcal{S}$, the probability to find a larger
value $\mathcal{S}' > \mathcal{S}$. This definition follows, as that
for the compatibility $\mathcal{C}$ in
Section~\ref{sec_gamma_short_overview}, the scheme given
in \citet{NeymanPearson1933} to construct a most efficient
hypothesis test.

For convenience, again the negative decadic logarithm is used instead,
\begin{align}
  \mathcal{T}(\mathcal{S}) = -\log_{10}
  \int_{\mathcal{S}}^{\infty}\mathrm{d}{s}\, f(\mathcal{S})\, ,
  \label{eq_gamma_def_T}
\end{align}
and the null hypothesis is rejected if $\mathcal{T} > 6.2$, which
corresponds to the common $5\sigma$
criterion. Figure~\ref{fig_gamma_epdf} plots
Eq.~\eqref{eq_gamma_def_T} for different system sizes and background
intensities, based on $10^9$ simulated count maps for each system.

As expected, the test statistic strongly depends on the background
intensity $\lambda$ because the number of thresholds with nonzero
deviation strength varies. Interestingly, the dependence on the system
size is rather weak.

The new test statistic is chosen such that in simulations of the
background model, its EPDF is the same as for the deviation strength
$\mathcal{D}$ at a single threshold,
\begin{align}
  f(\mathcal{T}) = \ln{10}\cdot{10}^{-\mathcal{T}}\;,
  \label{eq_gamma_distribution_of_T}
\end{align}
see Fig.~\ref{fig_gamma_distribution_of_T}.

\subsection{Sensitivity increase for diffuse radiation}

Especially for broad sources, which exhibit structural deviations at a
large range of thresholds, the new test statistic
$\mathcal{T}$ can lead to an additional increase in sensitivity.

\begin{figure}[t]
  \centering
  \includegraphics[width=\linewidth]{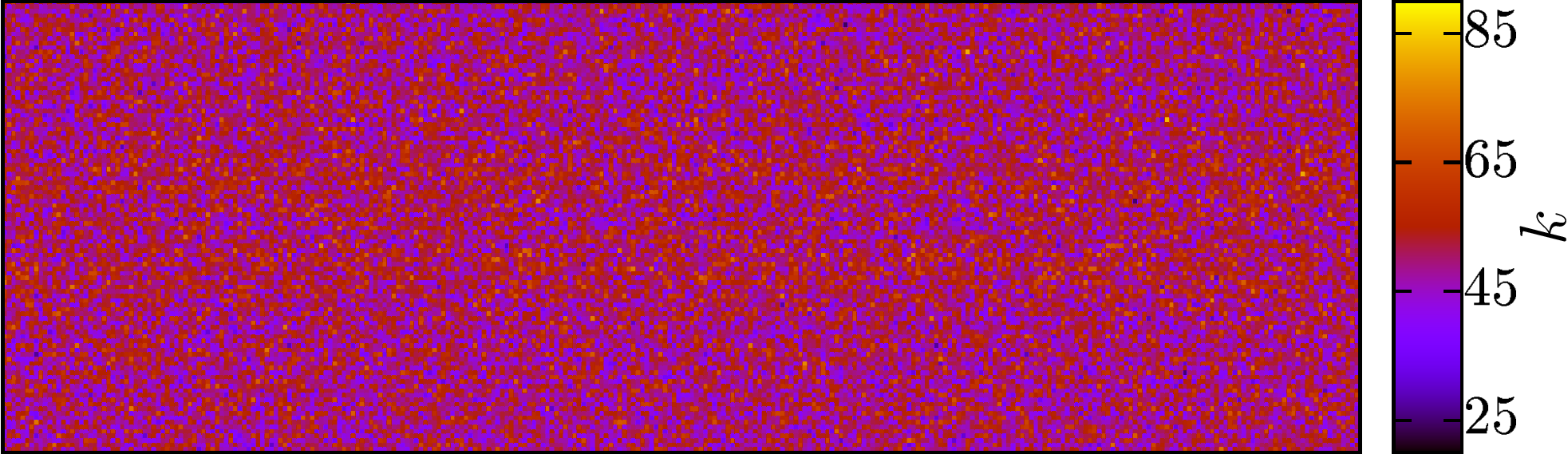}\\
  \includegraphics[width=\linewidth]{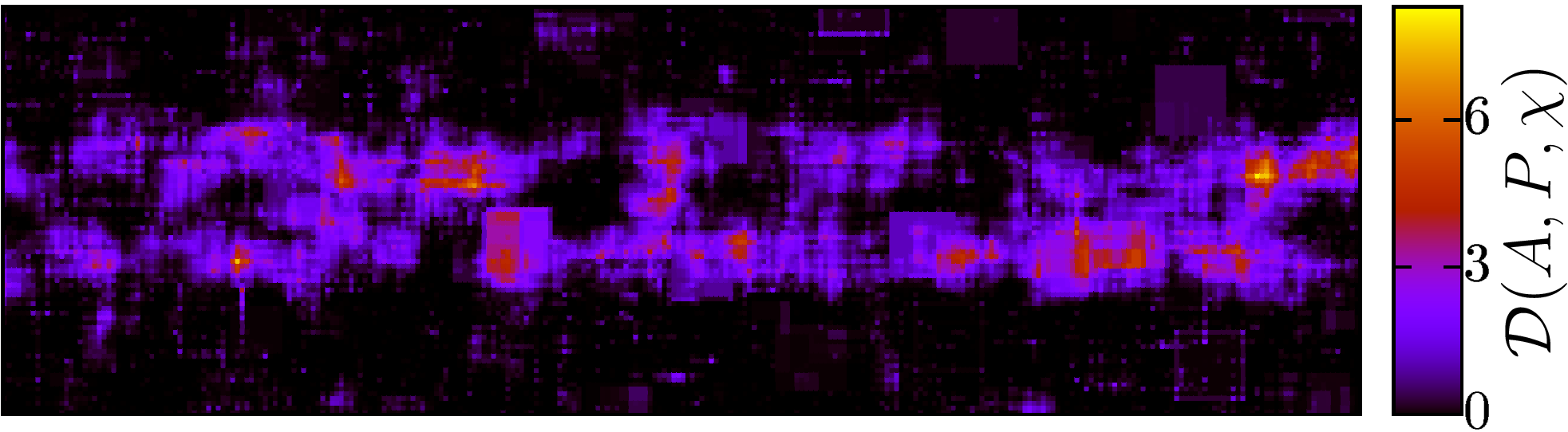}\\
  \includegraphics[width=\linewidth]{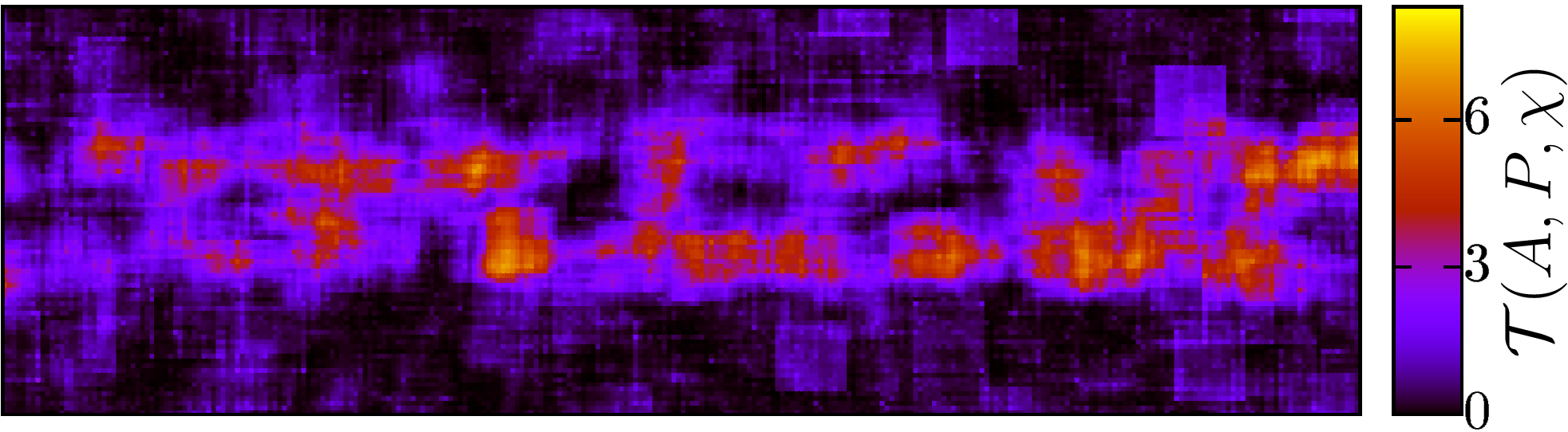}
  \caption{Diffuse radiation along the horizontal axis is added to a
    homogeneous background noise with intensity $\lambda=50$, which
    leads to an excess in the number of counts $k$ (top). The
    Minkowski sky map based on the maximum of the deviation strength
    $\mathcal{D}$ w.r.t. all Minkowski functionals detects the
    structural deviations of this very broad source (center), but the
    new test statistic that sums over all thresholds leads to a
    significant increase in sensitivity (bottom).}
  \label{fig_gamma_compare_D_to_T}
\end{figure}

Figure~\ref{fig_gamma_compare_D_to_T} exemplarily shows this increase
for simulated diffusion radiation. The new testing procedure is more
sensitive because it includes the information that there are also
strong structural deviations at other thresholds than the maximum
deviation strength.

% style aa.bst
\bibliographystyle{aa}
\bibliography{gamma-III}

\begin{thebibliography}{44}
\expandafter\ifx\csname natexlab\endcsname\relax\def\natexlab#1{#1}\fi

\bibitem[{Aharonian {et~al.}(2006{\natexlab{a}})}]{rxj1713}
Aharonian, F. {et~al.} 2006{\natexlab{a}}, Astron. Astrophys., 449, 223

\bibitem[{Aharonian {et~al.}(2006{\natexlab{b}})}]{galcen}
Aharonian, F. {et~al.} 2006{\natexlab{b}}, Nature, 439, 695

\bibitem[{Aharonian {et~al.}(2007)}]{velajr}
Aharonian, F. {et~al.} 2007, Astrophys. J., 661, 236

\bibitem[{Arfelli {et~al.}(2000)}]{ArfelliEtAl2000}
Arfelli, F. {et~al.} 2000, Radiology, 215, 286

\bibitem[{Atwood {et~al.}(2009)}]{fermilat}
Atwood, W.~B. {et~al.} 2009, Astrophys. J., 697, 1071

\bibitem[{Borg {et~al.}(2005)Borg, Thirde, Ferryman, Fusier, Valentin, Bremond,
  \& Thonnat}]{BorgEtAl2005}
Borg, M., Thirde, D., Ferryman, J., {et~al.} 2005, in {{Advanced Video and
  Signal Based Surveillance, 2005. AVSS 2005. IEEE Conference on}}, 16

\bibitem[{Buckley {et~al.}(2008)Buckley, Byrum, Dingus, Falcone, Kaaret,
  Krawzcynski, Pohl, Vassiliev, \& Williams}]{BuckleyEtAl2008}
Buckley, J., Byrum, K., Dingus, B., {et~al.} 2008, ArXiv e-prints,
  arXiv:0810.0444

\bibitem[{Canuto {et~al.}(2009)Canuto, McLachlan, Kettunen, Velic, Krishnan,
  Neves, de~Backer, Hu, Hobson, \& Brindle}]{CanutoEtAl2009}
Canuto, H.~C., McLachlan, C., Kettunen, M.~I., {et~al.} 2009, Magn. Reson.
  Med., 61, 1218

\bibitem[{Colombi {et~al.}(2000)Colombi, Pogosyan, \& Souradeep}]{Colombi2000}
Colombi, S., Pogosyan, D., \& Souradeep, T. 2000, Phys. Rev. Lett., 85, 5515

\bibitem[{Ducout {et~al.}(2013)Ducout, Bouchet, Colombi, Pogosyan, \&
  Prunet}]{Ducout2013}
Ducout, A., Bouchet, F.~R., Colombi, S., Pogosyan, D., \& Prunet, S. 2013, Mon.
  Not. R. Astron. Soc., 429, 2104

\bibitem[{Gay {et~al.}(2012)Gay, Pichon, \& Pogosyan}]{Gay2012}
Gay, C., Pichon, C., \& Pogosyan, D. 2012, Phys. Rev. D, 85, 023011

\bibitem[{{G\"oring}(2008)}]{Goering2008}
{G\"oring}, D. 2008, Master's thesis ({Diplomarbeit}), Universit\"at
  Erlangen-N\"urnberg

\bibitem[{{G\"oring}(2012)}]{Goering2012}
{G\"oring}, D. 2012, {{G}amma-Ray Astronomy Data Analysis Framework based on
  the Quantification of Background Morphologies using Minkowski Tensors}, PhD
  thesis, Universit\"at Erlangen-N\"urnberg

\bibitem[{G\"oring {et~al.}(2013)G\"oring, Klatt, Stegmann, \&
  Mecke}]{GoeringKlattEtAl2013}
G\"oring, D., Klatt, M.~A., Stegmann, C., \& Mecke, K. 2013, Astron.
  Astrophys., 555, A38

\bibitem[{Jain {et~al.}(2000)Jain, Duin, \& Mao}]{JainEtAl2000}
Jain, A.~K., Duin, R. P.~W., \& Mao, J. 2000, IEEE T. Pattern Anal., 22, 4

\bibitem[{{Kerscher} {et~al.}(2001){Kerscher}, {Mecke}, {Schmalzing},
  {Beisbart}, {Buchert}, \& {Wagner}}]{KerscherMecke2001}
{Kerscher}, M., {Mecke}, K., {Schmalzing}, J., {et~al.} 2001, Astron.
  Astrophys., 373, 1

\bibitem[{{Klatt}(2010)}]{Klatt2010}
{Klatt}, M.~A. 2010, Master's thesis ({Diplomarbeit}), Universit\"at
  Erlangen-N\"urnberg

\bibitem[{Klatt(2016)}]{Klatt2016}
Klatt, M.~A. 2016, Morphometry of random spatial structures in physics, PhD
  thesis, Friedrich-Alexander-Universit\"at Erlangen-N\"urnberg (FAU).

\bibitem[{Klatt {et~al.}(2012)Klatt, G\"oring, Stegmann, \&
  Mecke}]{KlattEtAl2012}
Klatt, M.~A., G\"oring, D., Stegmann, C., \& Mecke, K. 2012, {AIP Conf. Proc.},
  1505, 737

\bibitem[{Klatt \& Torquato(2014)}]{KlattTorquato2014}
Klatt, M.~A. \& Torquato, S. 2014, {Phys. Rev. E}, 90, 052120

\bibitem[{Larkin {et~al.}(2014)Larkin, Canuto, Kettunen, Booth, Hu, Krishnan,
  Bohndiek, Neves, McLachlan, Hobson, \& Brindle}]{LarkinEtAl2014}
Larkin, T.~J., Canuto, H.~C., Kettunen, M.~I., {et~al.} 2014, Magn. Reson.
  Med., 71, 402

\bibitem[{{Li} \& {Ma}(1983)}]{LiMa1983}
{Li}, T.-P. \& {Ma}, Y.-Q. 1983, Astrophys. J., 272, 317

\bibitem[{{M. Kerscher} {et~al.}(2001){M. Kerscher}, {K. Mecke}, {P.
  Schuecker}, {H. Böhringer}, {L. Guzzo}, {C. A. Collins}, {S. Schindler}, {S.
  De Grandi}, \& {R. Cruddace}}]{KerscherEtAl2001}
{M. Kerscher}, {K. Mecke}, {P. Schuecker}, {et~al.} 2001, Astron. Astrophys.,
  377, 1

\bibitem[{Mantz {et~al.}(2008)Mantz, Jacobs, \& Mecke}]{MantzJacobsMecke2008}
Mantz, H., Jacobs, K., \& Mecke, K. 2008, J.~Stat.~Mech., 12, P12015

\bibitem[{Mattox {et~al.}(1996)}]{egretlike}
Mattox, J.~R. {et~al.} 1996, Astrophys. J., 461, 396

\bibitem[{Mecke {et~al.}(1994)Mecke, Buchert, \&
  Wagner}]{MeckeBuchertWagner1994}
Mecke, K., Buchert, T., \& Wagner, H. 1994, Astron. Astrophys., 288, 697

\bibitem[{Michel {et~al.}(2013)}]{MichelEtAl2013}
Michel, T. {et~al.} 2013, Phys. Med. Biol., 58, 2713

\bibitem[{Neyman \& Pearson(1933)}]{NeymanPearson1933}
Neyman, J. \& Pearson, E.~S. 1933, Phil. Trans. R. Soc. London, Ser. A, 231,
  289

\bibitem[{Novaes {et~al.}(2014)Novaes, Bernui, Ferreira, \&
  Wuensche}]{NovaesEtAl2014}
Novaes, C., Bernui, A., Ferreira, I., \& Wuensche, C. 2014, J. Cosmol.
  Astropart. Phys., 2014, 018

\bibitem[{Novikov {et~al.}(2000)Novikov, Schmalzing, \&
  Mukhanov}]{Novikov:441276}
Novikov, D., Schmalzing, J., \& Mukhanov, V.~F. 2000, Astron. Astrophys., 364,
  17

\bibitem[{Quast \& Kaup(2011)}]{QuastKaup2011}
Quast, K. \& Kaup, A. 2011, EURASIP J. Image Video Process., 2011, 14

\bibitem[{Robins(2002)}]{Robins2002}
Robins, V. 2002, in Lecture Notes in Physics, Vol. 600, {Morphology of
  Condensed Matter}, ed. K.~Mecke \& D.~Stoyan (Springer, Berlin, Heidelberg),
  261

\bibitem[{Schmalzing {et~al.}(1999)Schmalzing, Buchert, Melott, Sahni,
  Sathyaprakash, \& Shandarin}]{Schmalzing1999}
Schmalzing, J., Buchert, T., Melott, A.~L., {et~al.} 1999, Astrophys. J., 526,
  568

\bibitem[{Schneider \& Weil(2008)}]{SchneiderWeil2008}
Schneider, R. \& Weil, W. 2008, {Stochastic and Integral Geometry (Probability
  and Its Applications)} (Berlin: Springer)

\bibitem[{Schr\"oder-Turk {et~al.}(2010)Schr\"oder-Turk, Kapfer, Breidenbach,
  Beisbart, \& Mecke}]{SchroederTurketal:2010jom}
Schr\"oder-Turk, G.~E., Kapfer, S., Breidenbach, B., Beisbart, C., \& Mecke, K.
  2010, J. Microsc., 238, 57

\bibitem[{Schr\"oder-Turk {et~al.}(2011)Schr\"oder-Turk, Mickel, Kapfer, Klatt,
  Schaller, Hoffmann, Kleppmann, Armstrong, Inayat, Hug, Reichelsdorfer,
  Peukert, Schwieger, \& Mecke}]{SchroederTurketal2010AdvMater}
Schr\"oder-Turk, G.~E., Mickel, W., Kapfer, S.~C., {et~al.} 2011, {Adv.
  Mater.}, 23, 2535

\bibitem[{Schr\"oder-Turk {et~al.}(2013)Schr\"oder-Turk, Mickel, Kapfer,
  Schaller, Breidenbach, Hug, \& Mecke}]{SchroederTurkEtAl2013NewJPhys}
Schr\"oder-Turk, G.~E., Mickel, W., Kapfer, S.~C., {et~al.} 2013, New J. Phys.,
  15, 083028

\bibitem[{Schuetrumpf {et~al.}(2013)Schuetrumpf, Klatt, Iida, Maruhn, Mecke, \&
  Reinhard}]{SchuetrumpfKlattEtAl2013}
Schuetrumpf, B., Klatt, M.~A., Iida, K., {et~al.} 2013, Phys. Rev. C, 87,
  055805

\bibitem[{Schuetrumpf {et~al.}(2015)Schuetrumpf, Klatt, Iida, Schr\"oder-Turk,
  Maruhn, Mecke, \& Reinhard}]{SchuetrumpfKlattEtAl2014}
Schuetrumpf, B., Klatt, M.~A., Iida, K., {et~al.} 2015, Phys. Rev. C, 91,
  025801

\bibitem[{Stonebraker {et~al.}(1993)Stonebraker, Frew, Gardels, \&
  Meredith}]{StonebrakerEtAl1993}
Stonebraker, M., Frew, J., Gardels, K., \& Meredith, J. 1993, SIGMOD Rec., 22,
  2

\bibitem[{Stoyan {et~al.}(1987)Stoyan, Kendall, \& Mecke}]{Stoyan1987}
Stoyan, D., Kendall, W., \& Mecke, J. 1987, {Stochastic geometry and its
  applications} (John Wiley and Sons)

\bibitem[{Theodoridis \& Koutroumbas(2009)}]{TheodoridisKoutroumbas2009}
Theodoridis, S. \& Koutroumbas, K. 2009, {Pattern Recognition}, {4th} edn.
  (Boston: {Academic Press})

\bibitem[{Wiegand {et~al.}(2014)Wiegand, Buchert, \&
  Ostermann}]{wiegand_direct_2014}
Wiegand, A., Buchert, T., \& Ostermann, M. 2014, Mon. Not. R. Astron. Soc.,
  443, 241

\bibitem[{Yilmaz {et~al.}(2006)Yilmaz, Javed, \& Shah}]{YilmazEtAl2006}
Yilmaz, A., Javed, O., \& Shah, M. 2006, {ACM} Comput. Surv., 38

\end{thebibliography}

\end{document}